# Large Scale Liquid Metal Batteries

V. Bojarevics, A. Tucs

University of Greenwich, London SE10 9LS, UK

E-mail: bv03@greenwich.ac.uk

**Abstract**

Liquid metal batteries are possible candidates for large scale energy storage offering a possible breakthrough of intermittent wind and solar energy exploitations. The major concern over their practical implementation is the operation at elevated temperatures and sensitivity to liquid motion. The concept of liquid metal battery bears a close similarity to aluminium electrolytic production cells. The two liquid layer MHD effects can be projected to the three liquid layer self-segregated structure of the batteries. This paper presents numerical models for the three density-stratified electrically conductive liquid layers using 3D and shallow layer approximation accounting for specific MHD effects during periods of battery activity. It is demonstrated that a stable operation of these batteries can be achieved if reusing an infrastructure of existing aluminium electrolysis pot lines. The basic principles of the MHD processes in the cells are illustrated by the numerical example cases.

**Introduction**

With the growing renewable energy production comes a requirement to store the intermittently produced electrical energy to be available at peak consumption times. The traditional battery storage is not well suited for a large scale capacity comparable to the needs of industrial or urban electrical grids. Liquid metal batteries are possible candidates for such large scale energy storage [1]. The attraction of liquid batteries lies in the fast kinetics at liquid metal-electrolyte interfaces, simple assembly and recycling, while the major difficulties to implementation are their sensitivity to liquid motion and operation at elevated temperatures [2]. Important features of the liquid only cells are the continuous renewal of the liquid metal electrodes upon charge-discharge cycling, as well as their insensitivity to micro-structural degradation mechanisms, which are a major limitation to the majority of solid battery types [3,4]. The concept of liquid metal battery bears a close similarity to aluminium electrolytic production cells containing two horizontal liquid layers, for which reliable mathematical models are available, see for instance [5] and the references therein. The two liquid layer MHD effects can be easily extended to the three liquid layer self-segregated structure of the batteries. The modern trend in gaining efficiency for commercial aluminium electrolysis cells is to increase their size instead of operating a large number of parallel small ones. We assume the same tendency will held for the efficient and practical implementation of large scale liquid metal battery cells having a large horizontal surface area and a small depth of the liquid metal electrodes and the electrolyte (which is the major electrical resistivity in the battery). The operation of the batteries needs to clarify if the layered structure will be overly sensitive to liquid motion caused mainly by electromagnetic forces, which potentially may lead to the top and bottom liquid metal electrode short-circuiting, accompanied by associated electrochemical and thermal problems [2]. The instability could result from interaction of the electric currents with the total magnetic field from all the current supply, connection lines and the self-induced magnetic field of the current in the cell. In order to prevent the most dangerous electromagnetically driven long wave instabilities [5,6], the electrolyte layer thickness should be above a certain critical limit, however the maximum thickness of electrolyte layer is limited by the requirement that the voltage loss in the electrolyte must not exceed a significant portion of the available thermodynamic driving force (open circuit voltage) [1]. For the relevant combination of metals and salts, this means that the electrolyte thickness should not exceed a few centimetres, while the horizontal extent of the liquid layers may reach several meters.

This paper presents numerical models for the three density-stratified electrically conductive liquid layers using both 3D and shallow layer approximation to account for specific MHD effects during periods of battery charging and discharging cycles. After considering a simple and straightforward battery design we arrive to the conclusion that a more complex magnetic stabilisation is required to make the batteries suitable for practical implementation. The knowledge accumulated in designing magnetically stable aluminium production cells [5,6] permits to investigate a possibility of reusing the infrastructure of an existing aluminium electrolysis pot-line. Such 'redundant' pot-lines are increasingly available due to the present overproduction of aluminium raw materials. The mathematical model for a typical aluminium electrolysis line of 180 kA cells is extended to the battery 3 liquid layer case while preserving the plant infrastructure arrangements. It is demonstrated that a reasonably stable operation of charging and discharging of these batteries can be achieved.

## 1. Problem set up and governing equations

The battery consists of the three liquid electrically conducting layers naturally stratified due to their density (see the Figure 1 for the notation to be used in this paper). The electric current from an external circuit is flowing upwards or downwards depending on the charge/discharge cycle. The liquid electrodes are highly conductive molten metals (about $10^6$ S/m), while the molten salt electrolyte has the conductivity of up to 300 S/m. The electric current enters and exits the cell via the metallic collectors connected to the external circuit. The current distribution in the cell is fully 3 dimensional. The problem can be solved by means of spectral algorithm based on Fourier and Chebyshev polynomial expansions:

$$\mathbf{j} = -\sigma_i \nabla \varphi, \quad \Delta \varphi_i = 0; \ x \in (0, L_x), y \in (0, L_y), z \in (H_{i-1}, H_i), \ i = 1, 2, 3 \quad (1)$$

$$\varphi_i(x, y, z) = \sum_{m,n=0}^{M,N} \sum_{p=0}^{P} a_{mnp}^{(i)} T_p(z) \cos(m\pi x / L_x) \cos(n\pi y / L_y). \quad (2)$$

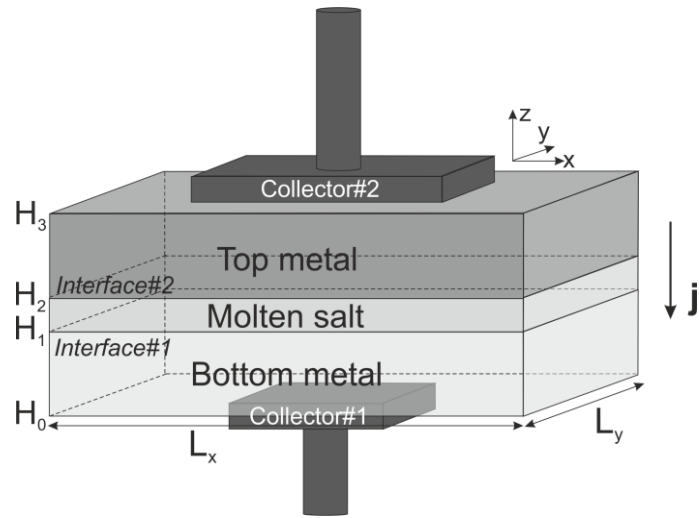

Figure 1. Schematic representation and the notation used for the liquid metal battery model.

The advantage of using the trigonometric expansion in the horizontal coordinate functions in (2) is that it satisfies the condition of insulating side-walls, while the given collector currents are imposed as the boundary conditions at the top and bottom electrode areas at $z=H_0$ and $z=H_3$. The interface conditions between the layers are the continuity of the normal electric current and the electric potential at the deformed liquid surfaces. The magnetic field (from the currents in the cell and the supply lines) is computed numerically using the Biot-Savart law for the volume current distribution in the cell and in the external circuit [5]:

$$\mathbf{B} = \frac{\mu_0}{4\pi} \int \frac{\mathbf{j} \times \mathbf{R}}{R^3} dV. \qquad (3)$$

The respective electromagnetic force distribution $\mathbf{f}(x,y,z,t) = \mathbf{j} \times \mathbf{B}$ is recomputed at each time step due to the electric current change in time with the interface $H_1(x,y,t)$ and $H_2(x,y,t)$ variation according to the hydrodynamic wave model. The wave model for the three coupled layers can be represented efficiently using the assumption of a small wave amplitude $A$ and the shallow layer approximation. The following non-dimensional parameters are introduced:

$$\varepsilon = A/h_1 \ll 1, \quad \delta = h_1/L_x, \quad E_k = IB_0 / \left( L_x^2 \rho_k g \varepsilon \delta \right), \qquad (4)$$

where the depth $h_1 = H_1 - H_0$ and the electromagnetic interaction parameter $E_k$ is scaled by the total current $I$ and the typical magnetic field $B_0$, relative to the gravitational restoring forces $\rho g$ arising for the small amplitude waves in a shallow layer. In the shallow layer approximation the effects of hydrodynamic friction at the bottom and top solid walls are taken into account by the terms with the coefficients $k_{f1}$ and $k_{f3}$ in the dynamic wave equations for the two coupled interfaces [5]:

$$\left(\frac{\rho_1}{h_1} + \frac{\rho_2}{h_2}\right)\partial_{tt}H_1 + \frac{\rho_2 k_{f1}}{h_1}\partial_t H_1 - \frac{\rho_2}{h_2}\partial_{tt}H_2 = (\rho_1 - \rho_2)g\partial_{jj}H_1$$
$$-E_1\partial_i F_{i1} + E_2\partial_i F_{i2} - \varepsilon\left[\rho_1\partial_j\left(U_{k1}\partial_k U_{j1}\right) - \rho_2\partial_j\left(U_{k2}\partial_k U_{j2}\right)\right], \qquad (5)$$

$$\left(\frac{\rho_2}{h_2} + \frac{\rho_3}{h_3}\right)\partial_{tt}H_2 + \frac{\rho_3 k_{f3}}{h_3}\partial_t H_2 - \frac{\rho_2}{h_2}\partial_{tt}H_1 = (\rho_2 - \rho_3)\partial_{jj}H_2$$
$$-E_2\partial_i F_{i2} + E_3\partial_i F_{i3} - \varepsilon\left[\rho_2\partial_j\left(U_{k2}\partial_k U_{j2}\right) - \rho_3\partial_j\left(U_{k3}\partial_k U_{j3}\right)\right]. \qquad (6)$$

The boundary conditions for the waves are derived from the zero normal velocity condition at the side walls:

$$\partial_n H_1 = (F_{n1} - F_{n2})/\left[(\rho_1 - \rho_2)g\right], \qquad (7)$$

$$\partial_n H_2 = (F_{n2} - F_{n3}) / [(\rho_2 - \rho_3)g]. \tag{8}$$

The horizontal depth average force $\mathbf{F}_k(x,y,t)$ and the velocity components $U_k$ (k=1,2) appearing in the wave equations are computed after the depth averaging procedure [5] is applied to the horizontal Navier-Stokes equation in the shallow layer approximation. The fully coupled MHD solution for the waves and fluid flow is validated extensively for the case of a single interface in the aluminium electrolysis cells. The solution is compared to measurements of the velocity fields [7,8] and the wave frequencies [6]. The computed and measured wave frequencies are similar to the typical gravity wave frequencies, however the exact values are shifted due to the MHD interaction. The numerically predicted frequency values are compared to the measurements on real industrial cells for various cases of disruption in the cell normal operation conditions using the wireless sensor technique [6].

## 2. Numerical solutions for the batteries

Computed results of the time dependent flow and interface wave development are presented for the case of a particular selection of the liquid metals: magnesium for the top 'light' layer of 0.2 m thickness, and the liquid antimony layer of similar thickness at bottom. This type of battery is relatively safe to realize in laboratory, and was used successfully in the small scale experiments [3]. More efficient (higher value EMF) combinations of the metals are available [3,4], however the Mg metal appears to be a challenging case due to the small density difference to the liquid electrolyte. The electrolyte density $\rho_2$= 1715 kg/m$^3$ in this case is close to the liquid magnesium density $\rho_3$= 1584 kg/m$^3$, making this a critical challenge to stabilize the wave motion after a small amplitude $A$=0.005 m gravity wave of the leading longitudinal mode (1,0) perturbation is applied as the initial disturbance. The surface of the bottom metal layer, composed initially of pure liquid antimony Sb, is expected to be more stable due to the higher density $\rho_1$= 6483 kg/m$^3$. The cell of the horizontal length 8 m by 3.6 m in width was used in the numerical experiments, which correspond to the dimensions of a commercial aluminium electrolysis cell. The depth of liquid electrolyte mid-layer was varied between 0.02 m to 0.08 m in order to investigate the stability of the MHD wave.

*2.1 Single collector cell*

The first attempts of modelling used simple single electric collectors at the top and the bottom of cell (Figure 2). The electric current of magnitude $I = 100$ kA is supplied by a single large collector at the top and removed by a single collector at the exit side of the bottom. The computed 3D electric current distribution for this case is shown in the Figure 3, demonstrating the presence of a high density horizontal current in the bottom metallic electrode layer. Due to the low conductivity of the electrolyte the electric current flows almost purely vertically in this layer from the large top collector until the surface of the electrolyte remains flat.

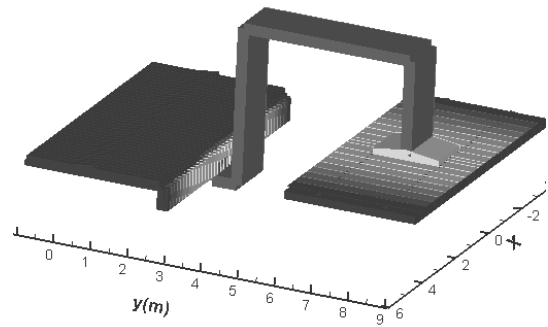

Fig. 2. The electrical connection between the single collector cells.

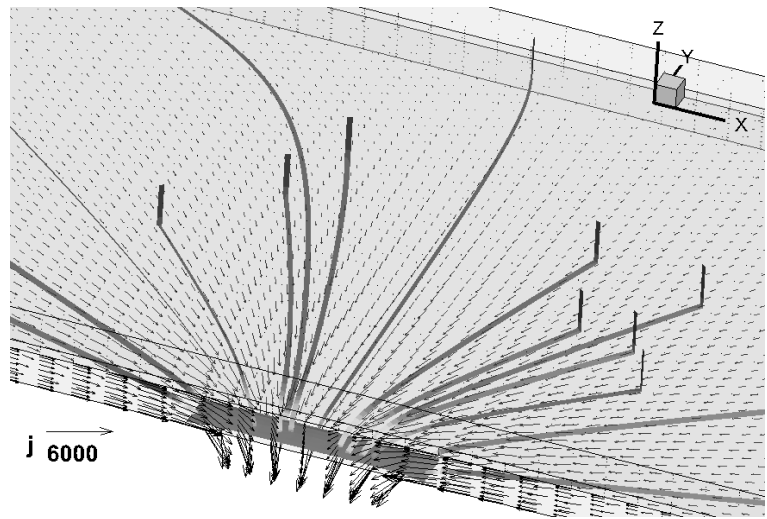

Fig. 3. The 3D solution for the electric current distribution in the liquid layers in the single collector battery.

The distribution of the total magnetic field in the liquid depends on the full path of the electric current including the surrounding cells and the return current line to the power source. The computed electromagnetic force is symmetric relative to the cell mid-plane if the return currents are unaccounted for. If the complete electrical circuit is included in the Biot-Savart law (3) integration volume, the magnetic field loses the symmetry. This leads the significant difference in the predicted velocity. The non-symmetric horizontal velocity in the bottom metal is shown on the left of the Figure 4, while the right side shows the symmetric vortex pattern obtained if ignoring the return current line. The velocity in the top metal and the electrolyte layer are different due to the differences in the electric current and the 3D distribution of the magnetic field.

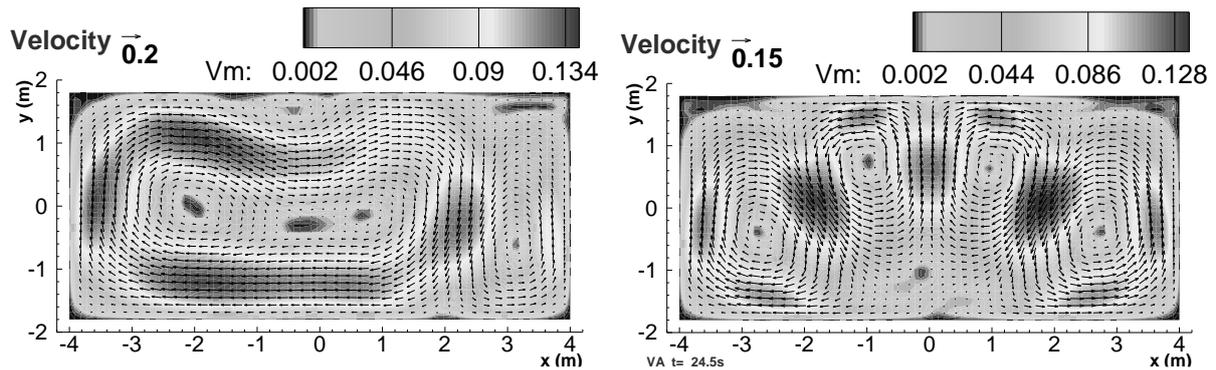

Fig. 4. The velocity field distribution in the bottom layer of the single collector battery at 75 kA current: the nonsymmetric pattern on the left for the full return current and the symmetric one at the right if ignoring the return line.

The long term simulation of the interface wave development with the fully coupled electromagnetic interaction demonstrates that the bottom heavier metal interface is stable to the perturbation effects. The top metal and the electrolyte interface is easily destabilized, due to the small density difference for the selected combination of materials, leading to the wave amplitude growth in a typical rotating wave instability [5]. The cell with the single current collectors can be stabilized only when reducing the magnitude of the total current $I$ from 100 kA to 75 kA, or alternatively increasing the electrolyte thickness from 0.05 m to 0.08 m. Figure 5 demonstrates the final interface shape after the initial perturbation as a longitudinal sloshing wave of amplitude A=0.005 m. The oscillation of the top metal interface is not completely damped even for the reduced current or the increased thickness, see the Figure 8, and continues to oscillate at the new frequency which is slightly shifted from the initial perturbation mode due to the MHD interaction. This battery could be possibly operated only at a lower current magnitude and at a low efficiency due to the high ohmic loss in the electrolyte. Figure 6 demonstrates the onset of the instability when the wave amplitude starts growing with the MHD interaction at the current value of 75 kA and low electrolyte thickness 0.05 m.

Remarkeably the bottom surface remains relatively flat during this process (Figure 6, right), apparently due to the significantly higher density of the bottom metal (Sb).

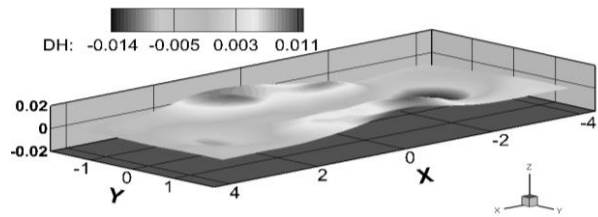

Fig. 5. The top interface shape at 75 kA and 0.08 m electrolyte thickness in the single collector battery.

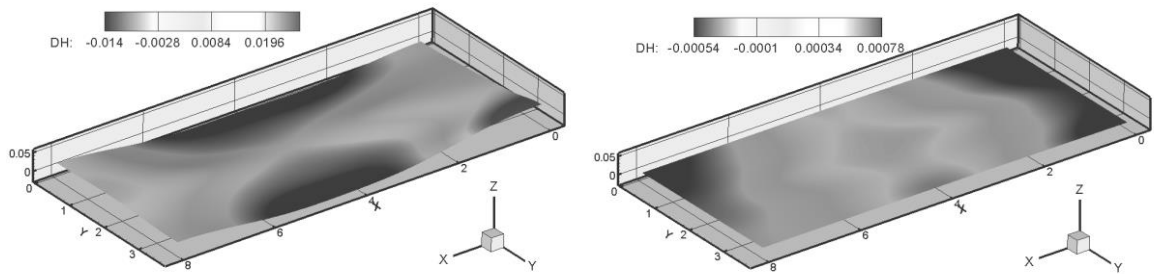

Fig. 6. The top (left) and botom (right) interface shape shortly before the electrolyte touches the top collector surface at 75 kA and 0.05 m electrolyte thickness in the single collector battery.

*Multiple collector battery cell*

An improvement in efficiency (by reducing the thickness of the electrolyte layer) can be achieved by optimizing the current supply path in such a way that the vertical magnetic field is reduced in magnitude and its distribution permits to avoid the MHD wave instability. The commercial aluminium electrolysis cells were developed following such stringent guidelines. Hence, we attempted to reuse one of the existing optimized bus bar configurations for the case of 3 liquid layers filling the cell cavity. The configuration of the current collectors, supply bars and the cell arrangement in rows is shown in the Figure 7. The multiple collectors ensures a uniform supply of the electric current to the top layer. In the liquid layers it has reduced horizontal components, particularly in the longer dimension 'x' of the cell, see the Figure 8. The corresponding computed magnetic field, particularly the $B_z$ component, is much smaller for this magnetically optimized cell at the same value of the total electric current $I = 100$ kA.

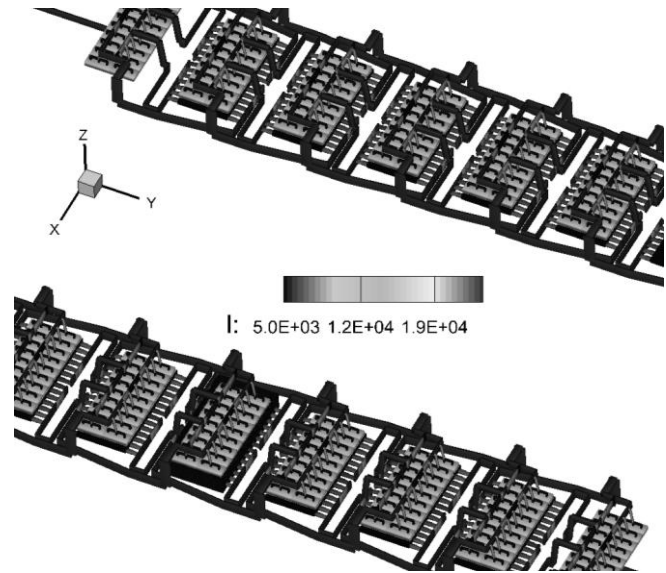

Figure 7. The configuration of the current collectors, supply bars and the arrangement of cells similar to the Trimet aluminium production cells.

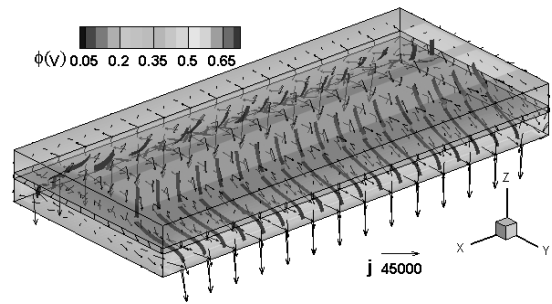

Figure 8. The 3D solution for the electric current distribution with the multiple current collectors.

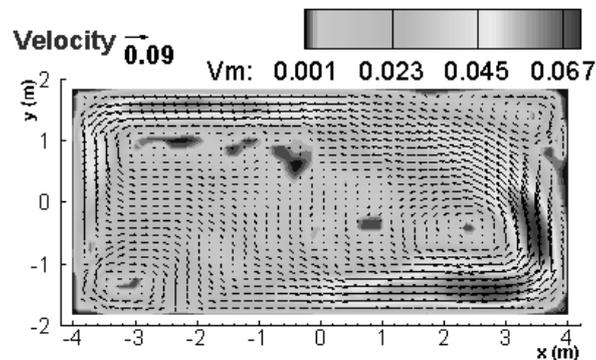

Figure 9. The velocity field distribution in the bottom layer of the multiple collector battery stabilised at 100 kA and 0.025 m electrolyte thickness.

The respective computed velocities are of smaller magnitude (Figure 9) if compared to the previous case, and the interface waves are stable after the initial perturbation dies out even at the full current of 100 kA, Figure 10. When the critical electrolyte thickness is reduced from 0.05 to 0.025 m, and further to the low value of 0.02 m. This means that it is possible to reach the stable discharge of this battery at a reasonable voltage drop of 0.29 V for the Mg/Bi metal combination (having the EDS = 0.4 V). The maximum current in the latter case is 75 kA. An additional advantage in this case is that the minimum thickness of the metal layers can be safely reduced to 0.1 m while maintaining the 3 layer battery completely stable in the MHD sense. After the initial perturbation is damped the interfaces are slightly deformed in a stationary shapes as shown in the Figure 11, however the deviation from the flat surface is just a couple of millimeters at most.

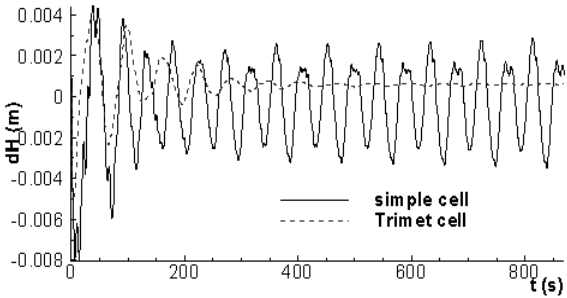

figure 10. The top metal oscillations for the single collector cell (100 kA, 0.08 m electrolyte) compared to the damped oscillation in the multiple collector type cell (100 kA, 0.025 m electrolyte).

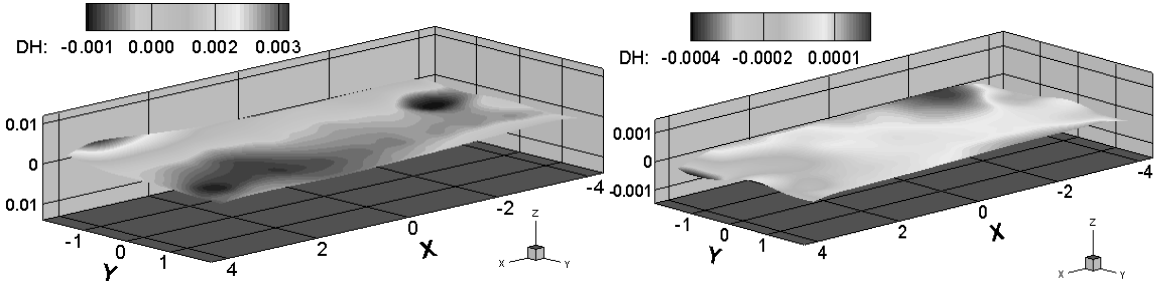

Figure 11. The top (left) and botom (right) interface shapes at 75 kA and 0.025 m electrolyte thickness in the multiple collector battery.

**Conclusions**

The MHD model for the large scale liquid metal battery demonstrates that it is possible to design a stable operating cell relative to dynamic perturbations if using the optimized bus bar configuration with multiple current collectors. For the Mg/Sb metal combination the bottom heavier metal layer is always stable to perturbations, leaving the top lighter metal (Mg, Li, Ca) interface stability as the critical step to control.